\documentclass[10pt]{article}
\usepackage[T1]{fontenc}
\usepackage{geometry}
\usepackage{graphics}
\usepackage{setspace}
\makeatletter

\begin{document}

{\par\centering {\Large Angular Correlation in Double Photoionization of Atoms
and the Role of the Observer }\Large \par}
\vspace{0.1in}

{\par\centering Dipankar Chattarji and Chiranjib Sur \par}

{\par\centering \emph{Department of Physics, Visva-Bharati, Santiniketan 731
235, INDIA}\par}
\vspace{0.3in}

{\small The problem of angular correlation in the double photoionization (DPI)
of rare gas atoms is considered in some depth. We refer particularly to the
efficiency operator for the detection of an electron by a detector having the
shape of a right circular cylinder. The different factors in the efficiency
operator are discussed in detail keeping in mind the fundamental epistemological
question of the role of the observer ( or his equipment) in such experiments.}{\small \par}
\vspace{0.2in}

\textbf{\small PACS No} {\small : 32.80.H, 32.80.F, 03.65.T,79.20.F }{\small \par}
\vspace{0.1in}

In this paper we wish to consider in some depth the problem of angular correlation
between the two electrons emitted by an atom when it is doubly ionized by a
photon. 

Consider a randomly oriented rare gas atom in a \( ^{1}S^{e} \) state. The
atom absorbs a photon and after a certain time interval emits a photo-electron
from one of the inner shells giving a singly ionized atomic state. This intermediate
ionic state now de-excites by emitting an Auger electron, typically from an
outer shell, giving rise to a two-vacancy final atomic state. We can denote
this sequence of events as follows.

\begin{equation}
\label{one}
h\nu +\mathbf{A}\longrightarrow \mathbf{A}^{+}+e_{1}^{-}\longrightarrow \mathbf{A}^{++}+e_{1}^{-}+e_{2}^{-}.
\end{equation}
 A polar plot of the observed distribution of coincidences between the two emitted
electrons as a function of the angle~ between their directions of emission shows
a clear periodic behaviour {[}\ref{schmidt}{]}. The question we ask is: what
is the origin of this angular correlation? Could it have anything to do with
the equipment used to detect the electrons? On the face of it, this last question
may not seem so obvious. It will, however, become clearer as we proceed with
our discussion. 

Double photoionization (DPI) occurs when an atomic target like the one described
above is irradiated with a monochromatic photon beam from an advanced light
source, e.g. a synchrotron. Along with single photoionization (PI), there may
be events in which~ two electrons are emitted by an atom in quick succession.
In case the time interval between the successive emission of the two electrons
is substantially longer than the time taken by the first electron to leave the
interaction zone, DPI may be regarded as a two-step process {[}\ref{we1}{]}.
In other words, the emission of the two electrons may be regarded as being clearly
separated in time. This in its turn will depend on the energy imparted to the
atom by the incident photon.

We wish to obtain an angular correlation function for the two emitted electrons
in terms of the angle between their directions of emission. We shall do this
by considering an ensemble of such atomic systems belonging to all possible
quantum mechanical states \( Q \). Each state \( Q \) is labeled by the total
angular momentum \( J \), its projection \( M \), and the remaining set of
quantum numbers \( \alpha  \).

The angular correlation function \( W(\theta ) \) for the two emitted electrons
is the probability that the angle between their directions of emission is \( \theta  \).
Evidently this is a statistical quantity, and \( W(\theta ) \) would have to
be the ensemble average of the above probability.

Now, how do we determine this probability? Hopefully, we let the atomic system
attain the final state given in Eq.(\ref{one}), we set up two detectors at
a suitable distance from the reaction zone with their axes making an angle \( \theta ^{\prime } \)
with each other, and we try to detect coincidences between the photo- and Auger
electrons. The number of coincidences we can hope to detect will depend on two
distinct factors.

(i) There is a certain probability for the atomic system to attain the final
state. This is described by the appropriate matrix element of the density or
statistical operator \( \rho  \) {[}\ref{fano},\ref{blum}{]}.

(ii) Even if the system goes over to the final state, because of the finite
size of the detecting equipment and other limiting factors, a coincidence event
may or may not be detected. There is thus a finite probability \( \epsilon \, (0\leq \epsilon \leq 1) \)
that the event will be detected. This probability is represented by the efficiency
operator \( \varepsilon  \). It will depend on the size, position and geometrical
configuration of the detecting equipment, but not its internal physical or chemical
nature provided that there is full absorption of an electron within the material
of the detector {[}\ref{rose}{]}. The emphasis is on those geometrical properties
of the detecting system which enable it to accurately determine the direction
of emission of an electron.

Obviously, \( W(\theta ) \) will be given by the joint probability of the formation
of the final state and its detection by the detecting equipment, i.e. by the
product of \( \rho  \) and \( \varepsilon  \). For a given state \( Q \)
this joint probability will be \( \rho \varepsilon  \) . The average probability
\( \overline{\varepsilon } \) for the ensemble will be given by the trace of
the product matrix. We write

\begin{equation}
\label{two}
\begin{array}{cc}
\overline{\varepsilon } & =\sum _{Q}\varepsilon _{Q}\left\langle Q\right| \rho \left| Q\right\rangle \\
 & =\sum _{Q}\varepsilon _{Q}\rho _{QQ}\\
 & =Tr(\varepsilon \rho )\\
 & =Tr(\rho \varepsilon )\, .
\end{array}
\end{equation}
 Here \( \varepsilon _{Q} \) is the efficiency or probability of detection
of the state described by the quantum numbers \( Q \), and \( \rho _{QQ} \)
the probability of the system being in the particular state \( Q \) {[}\ref{coester}{]}.
Both \( \varepsilon  \) and \( \rho  \) are tensor operators. 

Since the angular correlation function happens to be the trace of a matrix {[}\ref{coester},\ref{ferguson}{]},
it will be invariant under a unitary transformation in Hilbert space. 

Another property of the system arises from the random orientation of the rare
gas atoms. The electrons emitted by them are unpolarized and will, on the average,
have spherical symmetry. Spherical symmetry will also hold \emph{effectively}
if the detector is insensitive to polarization. Hence the angular correlation
function itself can depend only on scalar invariants formed of the unit momentum
vectors of the two emitted electrons \( \widehat{\mathbf{p}_{1}} \) and \( \widehat{\mathbf{p}_{2}} \).
These invariants are given by the scalar product of spherical tensors {[}\ref{satchler}{]},
as follows:

\begin{equation}
\label{three}
\begin{array}{cc}
\mathbf{C}_{k}(\widehat{\mathbf{p}_{1}})\cdot \mathbf{C}_{k}(\widehat{\mathbf{p}_{2}}) & =\sum _{m}C_{km}(\widehat{\mathbf{p}_{1}})C^{\star }_{km}(\widehat{\mathbf{p}_{2}})\\
 & =P_{k}(\widehat{\mathbf{p}_{1}}\cdot \widehat{\mathbf{p}_{2}})=P_{k}(cos\theta )\, .
\end{array}
\end{equation}
 In Eq.(\ref{three}) \( P_{k}(cos\theta ) \) is a Legendre polynomial. The
index \( k \) will be restricted to the allowed values of the resultant of
the angular momenta \( j_{1} \) and \( j_{2} \) of the two emitted electrons.
Out of these, odd values of \( k \) will drop out because they would give odd
parity.

Now, from the elements of statistical mechanics, we know that \( \overline{\varepsilon } \)
is the expectation value (or average value) of the efficiency operator \( \varepsilon  \)
{[}\ref{terharr}{]}. It needs to be pointed out that this expectation value is a function of \(\theta\). Thus the angular correlation function \( W(\theta ) \)
is, to within a multiplying factor, just the expectation value of the efficiency
operator for a given value of \(\theta\). To be more precise, it represents the angle-dependent factor in the
expectation value~ of the efficiency operator.

Now, based on physical considerations, can we find an expression for the efficiency
operator?

We begin by noting that the efficiency operator represents the attenuation of
the probability of detecting a particle (in our case it is an electron) caused
by certain geometrical properties of the detecting system. Some of these factors
were mentioned above. Let us now try to write down an expression for the efficiency
operator for an electron in terms of those factors. 

Obviously, a co-ordinate representation would be the most appropriate for the
discussion of these factors. But what kind of co-ordinate system shall we use?

Because of the spherical symmetry of the system as discussed above, we use spherical
polar co-ordinates. However, as soon as an emitted electron begins to interact
with a detector, the spherical symmetry is lost. This is because, regardless
of the type of counter used, the geometry is usually that of a right circular
cylinder. Hence as the unpolarized electrons enter the detection zone they acquire an axial
symmetry about the detector axis. This change in the symmetry of the system
as we go from the reaction zone to the detection zone will have some bearing
on our expression for the efficiency operator for unpolarized electrons.

In Fig.1, \( A_{1} \) is the direction of the axis of a cylindrical detector,
say a Geiger-Muller counter, of radius \( r \) and thickness \( t \) . The
base of the detector is placed at a distance \( h \) from the centre of the
reaction zone, \( h \) being parallel to \( A_{1} \). The angular width of
the detector as seen from the centre of the target is \( 2\gamma  \), where
\( tan\, \gamma =\frac{r}{h} \). On a rough estimate, \( h \) is usually about
10 cm. 

Let \( A_{1} \) represent the axis of the detector set up to detect the photo-electron,
and \( A_{2} \) the axis of the detector receiving the Auger electron. As shown
in Fig. 1, the angle between the two axes is \( \theta ^{\prime } \). The directions
\( D_{1} \) and \( D_{2} \) are the directions of emission of the photo- and
Auger electrons respectively, their azimuthal angles measured with respect to
the detector axes being \( \beta _{1} \) and \( \beta _{2} \).

We can now go back to the form of the efficiency operator corresponding to a
single detector detecting an electron. It is a tensor operator of rank \( n \)
with \( (2n+1) \) components. We assert that a reduced matrix element of the
component labeled by \( \nu  \) of the tensor is given by 

\begin{equation}
\label{four}
\varepsilon _{n\nu }(jj^{\prime })=\sum _{\nu ^{\prime }}z_{n}c_{n\nu ^{\prime }}(jj^{\prime })D_{\nu \nu ^{\prime }}^{n}(\Re )\, .
\end{equation}
 Here \( \nu  \) and \( \nu ^{\prime } \) are projection quantum numbers
corresponding to the angular momentum \( n \). \( z_{n} \) is the attenuation
factor due to the finite size of a detector. It is different for different values
of \( n \). \( D_{\nu \nu ^{\prime }}^{n}(\Re ) \) is an element of
the rotation matrix for the three-dimensional rotation \( \Re =(\theta _{1}\theta _{2}\theta _{3}) \),
\( \theta _{1},\theta _{2},\theta _{3} \) being Euler angles. \( \Re  \) represents
an effective rotation of the observed direction of emission of the detected
electron due to the finite angular size of the detector. And the factor \( c_{n\nu ^{\prime }}(jj^{\prime }) \)
arises from the change of symmetry as the electron goes to the detection zone
from the reaction zone. 

Now let us try to understand the genesis of each of these factors.

(a) \emph{Attenuation due to absorption in a detector of finite size}. Let \( x(\beta ) \)
be the distance traversed in a detector by an electron incident on its base
at an angle \( \beta  \) with the axis. Let \( \tau  \) be the absorption
coefficient of the material inside the detector. Then the absorption will be
proportional to \( (1-exp(-\tau x(\beta ))) \). Since this depends on the~
azimuthal angle \( \beta  \), each term in the Legendre polynomial expansion
of the angular correlation function will get multiplied by a different attenuation
factor due to absorption.

Let us first consider the case of a single detector detecting, say, a photo-electron.
Here we can think in terms of an angular distribution measurement. The angular
distribution~ too can be written out as a Legendre polynomial expansion. The
attenuation factor multiplying \( P_{n}(cos\beta ) \) will be {[}\ref{rose}{]},

\begin{equation}
\label{five}
z_{n}=\frac{J_{n}}{J_{0}},
\end{equation}
 where

\begin{equation}
\label{six}
J_{n}=\int ^{\gamma }_{o}P_{n}(cos\beta )(1-exp(-\tau x(\beta )))sin\beta d\beta \, .
\end{equation}

For an angular correlation experiment with two detectors having finite size
attenuation factors \( z_{n}(1) \) and \( z_{n}(2) \) the total attenuation
factor for the \( n \)th term will be

\begin{equation}
\label{seven}
Z_{n}=z_{n}(1)z_{n}(2)\, .
\end{equation}

(b) \emph{Rotational attenuation due to finite angular size of the detector}.
Let \( D_{1} \) be the direction in which the photo-electron is emitted by
the rare gas atom. As shown in Fig.1, \( D_{1} \) makes an angle \( \beta _{1} \)with
the axis \( A_{1} \) of the detector. But, whatever be the value of \( \beta _{1} \),
the observer will take the direction of emission of the photo-electron to be
given by \( A_{1} \). In other words, there will be an effective rotation of
the direction of~ emission. Denoting this rotation by \( \Re _{1} \), the corresponding
factor in the expression for the efficiency operator~ of the detector receiving
the photo-electron will be \( D_{\nu \nu^{\prime}}^{n}(\Re _{1}) \).

For an angular correlation experiment such as ours, the directions of emission~
of the photo- and Auger electrons will undergo the effective rotations, say,
\( \Re _{1} \) and \( \Re _{2} \) respectively. The corresponding factor in
the expression for the resulting efficiency operator will be \( D_{\nu _{2}\nu _{1}}^{n}(\Re ^{-1}_{2}\Re _{1}) \) {[}\ref{we2}{]}.

( c ) \emph{Attenuation factor corresponding to the state of polarization}. We have already discussed
the axial symmetry acquired by unpolarized electrons as they enter the detection zone.
What does this do to their quantum mechanical state?

Let us first consider the semi-classical vector model. Axial symmetry about
the detector axis implies that the angular momentum vector of an electron can
only lie in the plane perpendicular to that axis, i.e. the \( xy \) plane.
Obviously, the \( z \)-component of its angular momentum will be zero. Now
going over to quantum mechanics, only those states will survive for which \( \nu =0 \). 
This calls for a projection operator having the form 

\begin{equation}
\label{ten}
c_{n\nu }(jj^{\prime })=N_{jj^{\prime }n}C^{jj^{\prime }n}_{\frac{1}{2}-\frac{1}{2}0}\delta _{\nu 0}\, ,
\end{equation}
 where \( N_{jj^{\prime }n} \) is a normalizing factor which turns out to be
\( \frac{\sqrt{2j+1}\sqrt{2j^{\prime }+1}}{4\pi }(-1)^{j-\frac{1}{2}+n} \)
{[}\ref{we2}{]}. In other words, only the factor \( c_{n0}(jj^{\prime }) \)
enters into our expression for the efficiency operator. 

A formal derivation of this result is given in our paper under reference {[}\ref{we2}{]}.
However, that derivation does not quite relate to the attenuation properties
of a detector. On the other hand, we feel that our present approach is physically
more transparent. It also has the virtue of throwing some light on a couple
of questions of fundamental epistemological interest. Does the observer (or
his equipment) have a role in this type of experiment? If so, what is that role
like? Obviously, such questions can be important from the standpoint of the theory of measurement.

Calculation of the angular correlation function can now go through as in reference
{[}\ref{we2}{]}. We finally get

\begin{equation}
\label{11}
\begin{array}{ccc}
W(\theta ) & = & \sum _{k}z_{k}(1)z_{k}(2)(-1)^{j_{1}+j_{2}}c_{k0}(j_{1}j^{\prime }_{1})c^{\star }_{k0}(j_{2}j^{\prime }_{2})\\
 &  & \times w(J_{b}J^{\prime }_{b}j_{1}j^{\prime }_{1},kJ_{a})w(J_{b}J^{\prime }_{b}j_{2}j^{\prime }_{2},kJ_{c})P_{k}(cos\theta )\, ,
\end{array}
\end{equation}
where the \( w \)'s are Racah coefficients, \( j_{1} \) and \( j_{2} \)
are angular momenta of the photo- and Auger electrons, \( J_{a},J_{b},J_{c} \)
the angular momenta of the atom in its initial, intermediate and final states
respectively. Here \( k \) is an even integer ranging from \( 0 \) to \( k_{max} \),
\( k_{max} \) being defined as follows. Let \( \left\{ \left\{ j_{1}+j^{\prime }_{1}\right\} _{max},\left\{ j_{2}+j^{\prime }_{2}\right\} _{max}\right\} _{min}=p \).
Then 

\[
\begin{array}{ccc}
k_{max} & = & p\, \, if\, p\, is\, even,\\
 & = & (p-1)\, if\, p\, is\, odd.
\end{array}\]
 The set of primed angular momentum quantum numbers represent virtual states which may arise from possible 
interaction with other atoms and
electrons. 

Our results for the double photoionization of xenon are discussed in reference
{[}\ref{we2}{]} in some detail.

One of the authors (CS) in indebted to the University Grants Commission of India
for support in the form of a junior research fellowship.

This paper is dedicated to the memory of Rabindranath Tagore, whose discussion
with Albert Einstein on the epistemology of science may be recalled in this
connection {[}\ref{marianoff}{]}.

{\par\centering \_\_\_\_\_\_\_\_\_\_\_\_\_\_\_\_ \par}

\begin{enumerate}
\item \label{schmidt}B. K\( \ddot{a} \)mmerling and V. Schmidt, J.Phys.B \textbf{26},
1141(1991)
\item \label{we1}D. Chattarji and C. Sur, J. of Electron Spect. and Rel. Phen. \textbf{114-116},
153(2001)
\item \label{fano}U. Fano, Phys. Rev. \textbf{90}, 577(1953)
\item \label{blum}K. Blum, Density Matrix Theory and Applications (Plenum Press, New York, 
1981)
\item \label{rose}M. E. Rose, Phys. Rev. \textbf{91}, 610(1953)
\item \label{coester}F. Coester and J. M. Jauch, Helv. Phys. Acta \textbf{26}, 3(1953)
\item \label{ferguson}A. J. Ferguson, Angular Correlation Methods in Gamma-ray Spectroscopy
(North-Holland, Amsterdam, 1965)
\item \label{satchler}D. M. Brink and G. R. Satchler, Angular Momentum, 2nd ed.,
55(Oxford University Press, Oxford, 1968)
\item \label{terharr}D. ter Haar, Elements of Statistical Mechanics, 150 (Holt, Reinhart
and Winston, New York, 1960)
\item \label{we2}D. Chattarji and C. Sur, submitted to Phys. Rev. A (2001)
\item \label{marianoff}Dmitri Marianoff, The New York Times Magazine, Aug. 10, 1930
\end{enumerate}
{\par\centering \_\_\_\_\_\_\_\_\_\_\_\_\_\_\_\_ \par}

\vspace{0.5001cm}
{\par\centering \resizebox*{3in}{4.2in}{\includegraphics{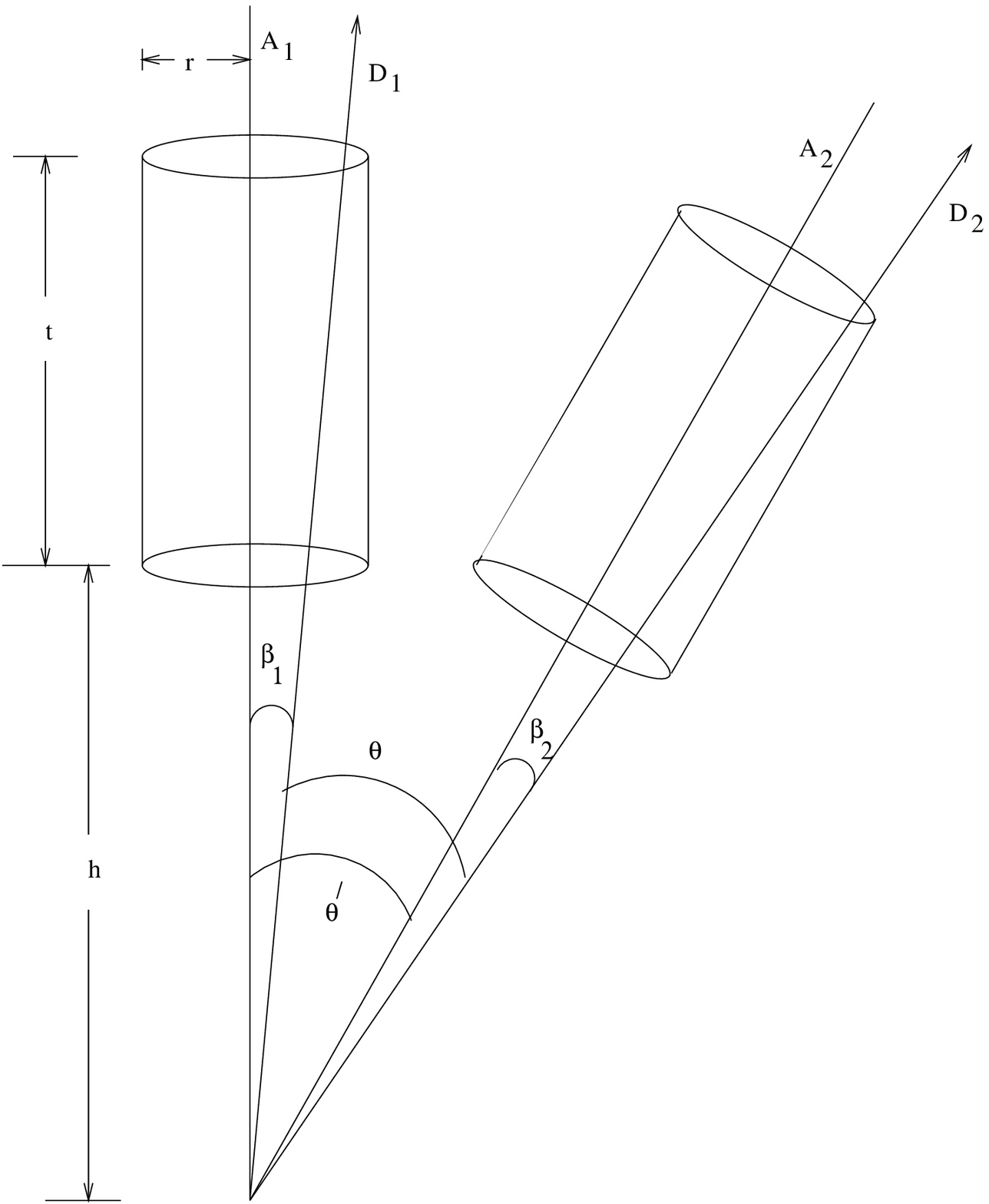}} \par}
\vspace{0.5001cm}

{\par\centering Fig 1 : Geometrical arrangement of the detectors\par}

\end{document}